\documentclass[prd,twocolumn,floatfix,superscriptaddress,nofootinbib]{revtex4}
\usepackage[utf8]{inputenc}
\usepackage{graphicx,epsfig}
\usepackage{amsmath}
\usepackage{amssymb}
\usepackage{multirow}
\usepackage{float}
\usepackage[usenames,dvipsnames]{color}
\usepackage[colorlinks]{hyperref}

\providecommand{\norm}[1]{\lVert#1\rVert}
\hypersetup{
     breaklinks=true,
     pdfstartview={FitH},
     colorlinks=true,
     linkcolor=blue,
     citecolor=red,
     filecolor=magenta,
     urlcolor=blue,
     anchorcolor=green,
     linktocpage=true
}

\begin{document}

\title{Binary Black Hole Coalescence and the Dynamics of Scalar Hair in Einstein--Maxwell--Scalar Theory}

\author{Jean Pierre D\'{i}az}
\email{jeanpierrerodney.diaz@userena.cl}
\affiliation{Departamento de F\'isica, Facultad de Ciencias, Universidad de La Serena,\\
Avenida Cisternas 1200, La Serena, Chile.}

\author{P. A. Gonz\'{a}lez}
\email{pablo.gonzalez@udp.cl}
\affiliation{Facultad de Ingenier\'{i}a y Ciencias, Universidad Diego Portales, Avenida Ej\'{e}rcito Libertador 441, Casilla 298-V, Santiago, Chile.}

\author{Eleftherios Papantonopoulos}
\email{lpapa@central.ntua.gr}
\affiliation{Physics Division, School of Applied Mathematical and Physical Sciences, National Technical University of Athens, 15780 Zografou Campus, Athens, Greece.}

\author{Yerko V\'asquez}
\email{yvasquez@userena.cl}
\affiliation{Departamento de F\'isica, Facultad de Ciencias, Universidad de La Serena,\\
Avenida Cisternas 1200, La Serena, Chile.}

\begin{abstract}
We investigate the head-on coalescence of charged binary black holes in Einstein--Maxwell--Scalar (EMS) theory using numerical relativity. The binaries are built from charged puncture initial data representing two Reissner--Nordstr\"om black holes immersed in a purely kinetic scalar perturbation: the scalar field initially vanishes, while its conjugate momentum provides a small seed for the instability. We evolve the coupled gravitational, electromagnetic, and scalar sectors and monitor the apparent horizons, the emitted radiation, and the scalar field on the horizons. Our simulations show that the nonminimal electromagnetic--scalar coupling can dynamically trigger the growth of scalar hair even when the individual black holes are initially scalar-free.  The subsequent evolution depends on the coupling parameter $\alpha_0$ and on the charge-to-mass ratio of the remnant.
For weak coupling, or when charge cancellation suppresses the electromagnetic source after the merger, the scalar field is radiated away or absorbed by the final horizon, and the system dynamically descalarizes. For sufficiently strong coupling and nonzero remnant charge, the scalar field remains finite, and the final black hole approaches a scalarized configuration.  For binaries with equal charge-to-mass ratios, $q_1=q_2=0.1$, our simulations identify a transition region between $1900<\alpha_0<2000$, that separates descalarized and scalarized remnants. The coalescence also excites scalar radiation whose time profile is qualitatively correlated with the dominant gravitational-wave mode during the nonlinear stage of the collision. These results provide a binary realization of scalarization/descalarization transitions in EMS theory and show that the fate of scalar hair is controlled by the interplay between the scalar coupling and the charge content of the remnant.
\end{abstract}

\maketitle
\tableofcontents

\section{Introduction}

The observation of gravitational waves (GWs) has opened a new window onto the
strong-field and highly dynamical regime of gravity. The first detection by the
LIGO--Virgo Collaboration provided direct evidence for the coalescence of a
binary black hole (BBH) system and marked the beginning of gravitational-wave
astronomy~\cite{LIGOScientific:2016aoc}. Since then, BBH mergers have become
one of the main laboratories for testing General Relativity (GR) in the
nonlinear regime. Most BBH simulations and waveform models assume vacuum
spacetimes, where the dynamics is governed only by the gravitational field.
However, in astrophysical, cosmological, and fundamental-physics contexts, it is
natural to ask how additional matter fields, effective charges, or new degrees
of freedom could modify compact-binary dynamics~\cite{Araya:2022few}. Charged
black holes and electromagnetic effects have already been explored in the
Einstein--Maxwell system~\cite{Moreno:2021neu,Palenzuela:2009hx}. An additional
extension is provided by Einstein--Maxwell--Scalar (EMS) theories, where the
electromagnetic sector is nonminimally coupled to a scalar field
~\cite{Astefanesei:2019pfq}. Such couplings are also motivated by related
dilaton and low-energy string-inspired models~\cite{Hirschmann:2017psw}, and
can give rise to electromagnetic and scalar radiation in addition to the usual
gravitational signal.

The information carried by gravitational waves encodes the properties of the
source, such as the component masses, the merger time, and the parameters of the
final black hole. Extracting this information requires comparing the observed
signal with waveform templates constructed within a given theoretical framework
~\cite{Flanagan:1997sx,Benjamin}. Post-Newtonian methods accurately describe
the early inspiral~\cite{Blanchet:2013haa}, but they cease to be reliable near
merger, where the dynamics become strongly nonlinear. This is the domain of
numerical relativity (NR), which provides the computational tools needed to solve
the Einstein equations and their extensions in the strong-field regime
~\cite{Alcubierre}. In practice, NR relies on suitable formulations of the
evolution problem, usually based on a separation of spacetime into spatial
hypersurfaces and time evolution. This \(3+1\) viewpoint goes back to the ADM
formulation and underlies many modern approaches~\cite{Arnowitt:1962hi,
Gourgoulhon:2007ue,Baumgarte}. Other formulations, such as the Generalized
Harmonic Coordinate system, have also been successfully used
~\cite{Lindblom:2005qh}, while \(3+1\)-based formulations remain standard in
many simulations of compact binaries and charged black holes
~\cite{Alcubierre:2009ij}.

The first successful long-term simulation of a BBH spacetime by
Pretorius~\cite{Pretorius:2005gq} initiated an era in which NR became an
essential tool for compact-object binaries. The vacuum BBH problem is now well
understood, while binaries involving additional matter fields or modified-gravity
degrees of freedom remain less explored. Recent developments include
semi-analytical BBH metric approximations tested with NR~\cite{Combi:2024inn},
constraints on ultralight bosons from BBH gravitational-wave observations
~\cite{Aswathi:2025nxa}, and studies of BBH dynamics, gravitational-wave
generation, and additional fields or dark-sector phenomenology
~\cite{Gliorio:2026yvh,Acevedo:2026xol,Mobilia:2026dxt}. In this broader
context, scalarization is particularly interesting: compact objects can acquire
a nontrivial scalar configuration, thereby evading the standard no-hair
expectations of GR.

Binary black hole mergers in scalar-tensor and curvature-coupled theories
provide an important point of comparison for the present work. NR simulations of
black hole binaries embedded in scalar-field environments showed that scalar
radiation can be generated during the late inspiral and merger, even when the
corresponding vacuum binary would be indistinguishable from its GR counterpart
~\cite{Healy:2011ef}. In Einstein--scalar--Gauss--Bonnet gravity, simulations
in the decoupling limit showed that binary mergers can either produce a
scalarized remnant or dynamically descalarize through the emission of scalar
radiation~\cite{Silva:2020omi}. Subsequent studies further demonstrated that
spin can trigger scalarization, descalarization, or stealth-like behavior during
black hole coalescence~\cite{Elley:2022ept}. These results show that scalar hair
in binary systems can be genuinely dynamical and motivate the study of analogous
mechanisms in EMS binaries, where the scalar field is sourced by the
electromagnetic sector rather than by curvature invariants.

Charged black holes are also a natural extension of the standard vacuum BBH
problem. Numerical techniques have been developed to construct and evolve
electrically charged black holes in the Einstein--Maxwell system
~\cite{Alcubierre:2009ij,Zilhao:2012gp}. Although astrophysical black holes are
usually expected to be nearly neutral in many environments, charged or
effectively charged configurations provide useful theoretical laboratories for
studying the interplay between gravitational and electromagnetic fields. They
are especially relevant in EMS theory because the scalar field couples directly
to electromagnetic invariants. More generally, scalarized black hole solutions
are known in several extensions of GR, including theories with scalar couplings
to the Gauss--Bonnet invariant~\cite{East:2022rqi}, Horndeski theories
~\cite{Figueras:2021abd}, EMS theories~\cite{Herdeiro:2018wub,
Astefanesei:2019pfq}, and teleparallel gravity frameworks
~\cite{Bahamonde:2022lvh,Bahamonde:2022chq,Gonzalez:2024ifp}. Analytical hairy charged black-hole solutions have also been obtained in Einstein--Maxwell--Scalar models with AdS and Lifshitz asymptotics~\cite{Bravo-Gaete:2025lgs, Bravo-Gaete:2026bdq}. Such models are
also connected to broader open questions in gravitational physics, including
dark matter, dark energy, extra dimensions, and quantum gravity. Exotic compact
objects in EMS theories have also been studied in this context~\cite{DeFelice:2026cse}.

In EMS theory, the existence and dynamics of scalar hair are controlled by the
choice of scalar--electromagnetic coupling. For the scalarized-connected class
of models, the scalar-free Reissner--Nordstr\"om branch remains a solution of
the theory but can become tachyonically unstable under scalar perturbations.
The nonlinear development of this instability has been studied for isolated
charged black holes: a small scalar seed can grow and drive the system toward a
scalarized black hole configuration~\cite{Xiong:2022ozw}. The complementary
process of dynamical descalarization has also been analyzed, showing that scalar
hair can be radiated away or absorbed by the horizon and that the effective
charge-to-mass ratio plays a central role in determining whether the final black
hole remains scalarized or relaxes toward a scalar-free configuration
~\cite{Niu:2022zlf}. More recently, critical phenomena associated with
dynamical scalarization and descalarization have been identified in EMS models,
revealing threshold behavior between bald and scalarized end states
~\cite{Jiang:2023yyn}.

These results raise a natural question: how does scalarization behave in a
binary, strongly nonlinear spacetime? In particular, it is not evident whether
scalar hair generated before or during the merger will survive in the final
black hole, or whether it will be lost through scalar radiation and absorption
after coalescence. In a binary EMS system, the final state is expected to depend
not only on the scalar coupling and on the individual black-hole charges, but
also on the charge retained by the merger remnant. This makes BBH coalescence in
EMS theory a useful setting for exploring dynamical scalarization,
descalarization, and scalar radiation beyond the isolated-black-hole problem.

In this work, we study the head-on coalescence of charged binary black holes in
Einstein--Maxwell--Scalar theory using numerical relativity. We construct
charged puncture initial data representing two Reissner--Nordstr\"om black holes
and evolve the coupled gravitational, electromagnetic, and scalar sectors with
GRChombo~\cite{Andrade:2021rbd}. The initial scalar field is taken to vanish,
while a small profile for its conjugate momentum provides a purely kinetic seed
for the scalar dynamics. This setup allows us to test whether the
electromagnetic--scalar coupling can dynamically generate scalar hair in a
binary spacetime and whether such hair survives after the merger. We find that the
scalar field can be amplified near the horizons during the infall and merger,
that the collision excites scalar radiation qualitatively correlated with the
dominant gravitational-wave mode, and that the remnant may either descalarize or
approach a scalarized configuration depending on the coupling strength and on
the charge retained after coalescence.

The paper is organized as follows. In Sec.~\ref{formulation} we introduce the
EMS theory, the scalarization mechanism, and the numerical evolution system used
in our simulations. In Sec.~\ref{initialdata} we describe the charged puncture
initial data, the purely kinetic scalar perturbation that seeds the scalar
dynamics, and the numerical setup. In Sec.~\ref{results} we present the results
for head-on charged BBH coalescences, including constraint monitoring,
dynamical scalarization and descalarization, radiation extraction, remnant
formation, and the comparison between equal-charge and opposite-charge binaries.
Finally, in Sec.~\ref{Conclusions} we summarize our conclusions and outline
future directions.

\section{EMS theory and numerical formulation}
\label{formulation}

\subsection{Action, field equations and scalarization mechanism}

We consider EMS theories described by the action
\begin{equation}
S=\int d^{4}x\sqrt{-g}\left[ R-2\nabla_{\mu}\phi\nabla^{\mu}\phi-f(\phi)F^{\mu\nu}F_{\mu\nu} \right] ,
\label{eq:action}
\end{equation}
where $F_{\mu\nu}=\partial_{\mu}A_{\nu}-\partial_{\nu}A_{\mu}$ is the Maxwell tensor and $f(\phi)$ is the scalar--electromagnetic coupling. The equations of motion for the electromagnetic and scalar sectors are
\begin{eqnarray}
\nabla_{\mu}F^{\mu\nu}&=&-\frac{\dot f(\phi)}{f(\phi)}F^{\mu\nu}\nabla_{\mu}\phi ,
\label{eq:maxwell}\\
\nabla_{\mu}\nabla^{\mu}\phi&=&-\frac{1}{4}\dot f(\phi)F^{\mu\nu}F_{\mu\nu} ,
\label{eq:scalar}
\end{eqnarray}
where a dot denotes differentiation with respect to $\phi$. The corresponding energy-momentum tensor is
\begin{equation}
T_{\mu\nu}=T^{\rm EM}_{\mu\nu}+f(\phi)T^{\phi}_{\mu\nu},
\end{equation}
with
\begin{eqnarray}
4\pi T^{\rm EM}_{\mu\nu}&=&F_{\mu\alpha}F_{\nu}^{\ \alpha}-\frac{1}{4}g_{\mu\nu}F_{\alpha\beta}F^{\alpha\beta},\\
4\pi T^{\phi}_{\mu\nu}&=&\nabla_{\mu}\phi\nabla_{\nu}\phi-\frac{1}{2}g_{\mu\nu}\nabla_{\alpha}\phi\nabla^{\alpha}\phi .
\end{eqnarray}

EMS theories include a variety of scalar--electromagnetic couplings. For
example, dilatonic-type couplings of the form \(f(\phi)=e^{-2a\phi}\) arise in
models inspired by dimensional reduction and low-energy string theory; the
choice \(a=\sqrt{3}\) is associated with Kaluza--Klein theory
~\cite{Kaluza:1921tu}, while other values are related to string-inspired
dilatonic black-hole models~\cite{Hirschmann:2017psw}. Closely related exponential couplings,
such as \(f(\phi)=e^{\alpha_0\phi^2}\), have been extensively used to study the
coupling dependence and dynamical features of spontaneous scalarization of
charged black holes~\cite{Fernandes:2019rez}. In this work, however, we focus on the scalarized-connected quadratic coupling
\begin{equation}
f(\phi)=1+\alpha_0\phi^2\,,
\label{eq:coupling}
\end{equation}
which allows the scalar-free Reissner--Nordstr\"om solution but can trigger a
tachyonic scalar instability for sufficiently large \(\alpha_0\). 
It belongs to the scalarized-connected class of EMS models: $f'(0)=0$, so the scalar-free Reissner--Nordstr\"om solution remains an exact solution, while $f''(0)=2\alpha_0\neq0$ allows the scalar-free branch to become unstable. Linearizing Eq.~\eqref{eq:scalar} around $\phi=0$ gives
\begin{equation}
\left(\Box-\mu_{\rm eff}^{2}\right)\delta\phi=0,
\qquad
\mu_{\rm eff}^{2}=\frac{1}{4}f''(0)F_{\mu\nu}F^{\mu\nu} .
\label{eq:mueff}
\end{equation}
For an initially electric-dominated configuration, $F_{\mu\nu}F^{\mu\nu}<0$. Therefore, sufficiently large positive values of $\alpha_0$ can make $\mu_{\rm eff}^{2}<0$ near the horizons and trigger the tachyonic growth of the scalar field. This is the mechanism underlying the dynamical scalarization studied below. 

\subsection{3+1 variables and evolution system}

We use a standard $3+1$ decomposition of the spacetime metric,
\begin{equation}
ds^2=-\alpha^2dt^2+\gamma_{ij}(dx^i+\beta^idt)(dx^j+\beta^jdt),
\end{equation}
where $\alpha$, $\beta^i$ and $\gamma_{ij}$ are the lapse, shift, and spatial metric. Although the ADM decomposition provides the geometric basis for the Cauchy
problem, direct ADM evolutions are not the preferred choice for long-term
numerical simulations due to hyperbolicity and stability issues
~\cite{Friedman:2004jr}. This motivated the BSSN reformulation of the Einstein
equations, developed by Baumgarte--Shapiro and Shibata--Nakamura
~\cite{Baumgarte:1998te,Shibata:1995we}. Here, we use the CCZ4 formulation
implemented in GRChombo to evolve the spacetime geometry ~\cite{Bernuzzi:2009ex,Alic:2011gg,Andrade:2021rbd}. The lapse and shift are evolved using the standard moving-puncture gauge conditions~\cite{Bona:1994dr,Alcubierre:2002kk}.

Introducing the electric and magnetic fields through
\begin{equation}
F_{\mu\nu}=n_{\mu}E_{\nu}-n_{\nu}E_{\mu}+\epsilon_{\mu\nu\alpha}B^{\alpha},
\end{equation}
where $n^{\mu}$ is the future-directed timelike unit normal to the spatial hypersurfaces, corresponding to the four-velocity of the Eulerian observers, and $\epsilon_{\mu\nu\alpha} = n^{\beta} \epsilon_{\beta \mu \nu\alpha} $ is the induced spatial Levi-Civita tensor. We evolve the electromagnetic fields with divergence-cleaning variables $\Psi$ and $\Phi$
\begin{eqnarray}
\nabla_{\mu}(F^{\mu\nu}+g^{\mu\nu}\Psi)&=&\lambda_1 n^{\nu}\Psi-\frac{\dot f(\phi)}{f(\phi)}F^{\mu\nu}\nabla_{\mu}\phi\,,\\
\nabla_{\mu}(\star F^{\mu\nu}+g^{\mu\nu}\Phi)&=&\lambda_2 n^{\nu}\Phi\,,
\end{eqnarray}
where $\lambda_{1}$ and $\lambda_{2}$ are constants that parameterize the damping. The scalar field is evolved through its conjugate momentum $\Pi$,
\begin{eqnarray}
(\partial_t-\mathcal{L}_{\beta})\phi &=&- \alpha \Pi,\\
\notag (\partial_t-\mathcal{L}_{\beta})\Pi &=& -D_i(\alpha D^i\phi)+\alpha K\Pi+\alpha \dot V(\phi)\\
&& +\frac{1}{2}\alpha \dot f(\phi)(B_iB^i-E_iE^i),
\end{eqnarray}
where $V(\phi)=0$ in the simulations. The electromagnetic evolution equations are
\begin{eqnarray}
\notag (\partial_t-\mathcal{L}_{\beta})E^i &=& \epsilon^{ijk}D_j(\alpha B_k)+\alpha K E^i-\alpha D^i\Psi\\
&&+\alpha\frac{\dot f(\phi)}{f(\phi)}\left(\epsilon^{ijk}D_j\phi B_k+\Pi E^i\right),\\
(\partial_t-\mathcal{L}_{\beta})\Psi &=& -\alpha\left[\frac{\dot f(\phi)}{f(\phi)}D_j\phi E^j+D_jE^j+\lambda_1\Psi\right],\\
(\partial_t-\mathcal{L}_{\beta})B^i &=& -\epsilon^{ijk}D_j(\alpha E_k)+\alpha K B^i-\alpha D^i\Phi,\\
(\partial_t-\mathcal{L}_{\beta})\Phi &=& -\alpha(D_jB^j+\lambda_2\Phi).
\end{eqnarray}
The matter projections entering the Einstein equations are
\begin{eqnarray}
4\pi\rho &=& \frac{1}{2}D_i\phi D^i\phi+\frac{1}{2}\Pi^2+\frac{1}{2}f(\phi)(B_iB^i+E_iE^i),\\
4\pi S_i &=& \Pi D_i\phi+f(\phi)\epsilon_{ijk}E^jB^k .
\end{eqnarray}
These terms explicitly show how the choice of coupling affects both the matter sources of the spacetime evolution and the scalar-field dynamics.

\subsection{Numerical implementation}

The simulations are performed with GRChombo~\cite{Andrade:2021rbd}, using adaptive mesh refinement, fourth-order finite differences, fourth-order Runge--Kutta time integration, and Kreiss--Oliger dissipation~\cite{Berger,Kreiss}. The apparent horizons are located with AHFinder, and the horizon quantities are computed using the isolated-horizon formalism~\cite{Dreyer:2002mx,Franca:2023bed}. Gravitational radiation is extracted from the Newman--Penrose scalar $\Psi_4$ and decomposed into spin-weighted spherical harmonics~\cite{Baker:2001sf}.

\section{Initial data and numerical setup}
\label{initialdata}

\subsection{Charged puncture data}

The binary initial data are constructed with a charged-puncture solver implemented in this work, based on the TwoPunctures method \cite{Ansorg:2004ds} and on the charged-black-hole formulation of Ref. \cite{Bozzola:2019aaw}. The implementation was further extended to include the EMS scalar sector. 
The spatial metric is written in conformally flat form,
\begin{equation}
\gamma_{ij}=\psi^4\delta_{ij} \,.
\end{equation}
We also impose maximal slicing, \(K=0\)~\cite{Estabrook:1973ue}, so that the
trace of the extrinsic curvature vanishes  and the momentum constraint can be
treated through the standard Bowen--York construction. The Reissner--Nordstr\"om conformal factor in isotropic coordinates motivates the binary ansatz
\begin{equation}
\psi=\sqrt{\left(1+\sum_{n=1}^{N}\frac{m_n}{2r_n}+u\right)^2-
\left(\sum_{n=1}^{N}\frac{Q_n}{2r_n}\right)^2 } .
\label{eq:chargedpsi}
\end{equation}
Here, $m_n$ and $Q_n$ are the puncture mass and electric charge parameters, $r_n$ is the coordinate distance to the $n$-th puncture, and $u$ is the regular correction solved numerically.
 We define the dimensionless charge-to-mass ratio of each black hole as $q_n \equiv Q_n /M_n^{\rm ADM}$, where $M_n^{\rm ADM}$ denotes the corresponding individual ADM mass.
Defining
\begin{equation}
\eta=\sum_{n=1}^{N}\frac{m_n}{2r_n},\qquad
\sigma=\sum_{n=1}^{N}\frac{Q_n}{2r_n},\qquad
\kappa=1+u+\eta,
\end{equation}
the Hamiltonian constraint becomes an elliptic equation for $u$,
\begin{eqnarray}
\notag && \kappa\nabla^2u+\partial_i\kappa\partial^i\kappa-
\partial_i\sigma\partial^i\sigma-
\partial_i\psi\partial^i\psi+
\frac{1}{8}\psi^{-6}\bar A_{ij}\bar A^{ij}\\
&& +2\pi\psi^{-2}\bar\rho=0 .
\label{HRic}
\end{eqnarray}
The conformal electric field is taken to be the superposition of the individual Coulomb fields
\begin{equation}
\bar E^i=\sum_{n=1}^{N}\frac{Q_n}{r_n^2}\frac{x_n^i}{r_n},
\end{equation}
while the initial magnetic field is set to zero. Since \(B^i=0\) initially, the electromagnetic contribution to the momentum
density vanishes, consistently with the use of the Bowen--York solution for the
momentum constraint~\cite{York:1973ia,Bowen:1980yu}. 

The electric charge modifies the relation between the bare puncture masses and the individual ADM masses. Following the standard puncture inversion argument used to compute individual
ADM masses~\cite{Tichy:2003zg,Ansorg:2004ds}, and extending it to the
Reissner--Nordstr\"om conformal factor, for a binary with coordinate separation $D$ we use
\begin{equation}
M_A^{\rm ADM}=m_A(1+u_A)+\sum_{B\neq A}\frac{m_A m_B}{2D}
-\sum_{B\neq A}\frac{Q_A Q_B}{2D} .
\label{eq:admcharged}
\end{equation}
Here, \(u_A\) denotes the regular correction of the conformal factor evaluated at
the puncture \(A\), while \(Q_A\) is the electric charge of that puncture. The last
term is important when comparing equal-charge and opposite-charge binaries
because the sign of \(Q_AQ_B\) changes the interaction contribution to the
individual ADM masses.

\subsection{Purely kinetic scalar seed}

\begin{figure}[H]
    \begin{center}
        \includegraphics[width=86mm]{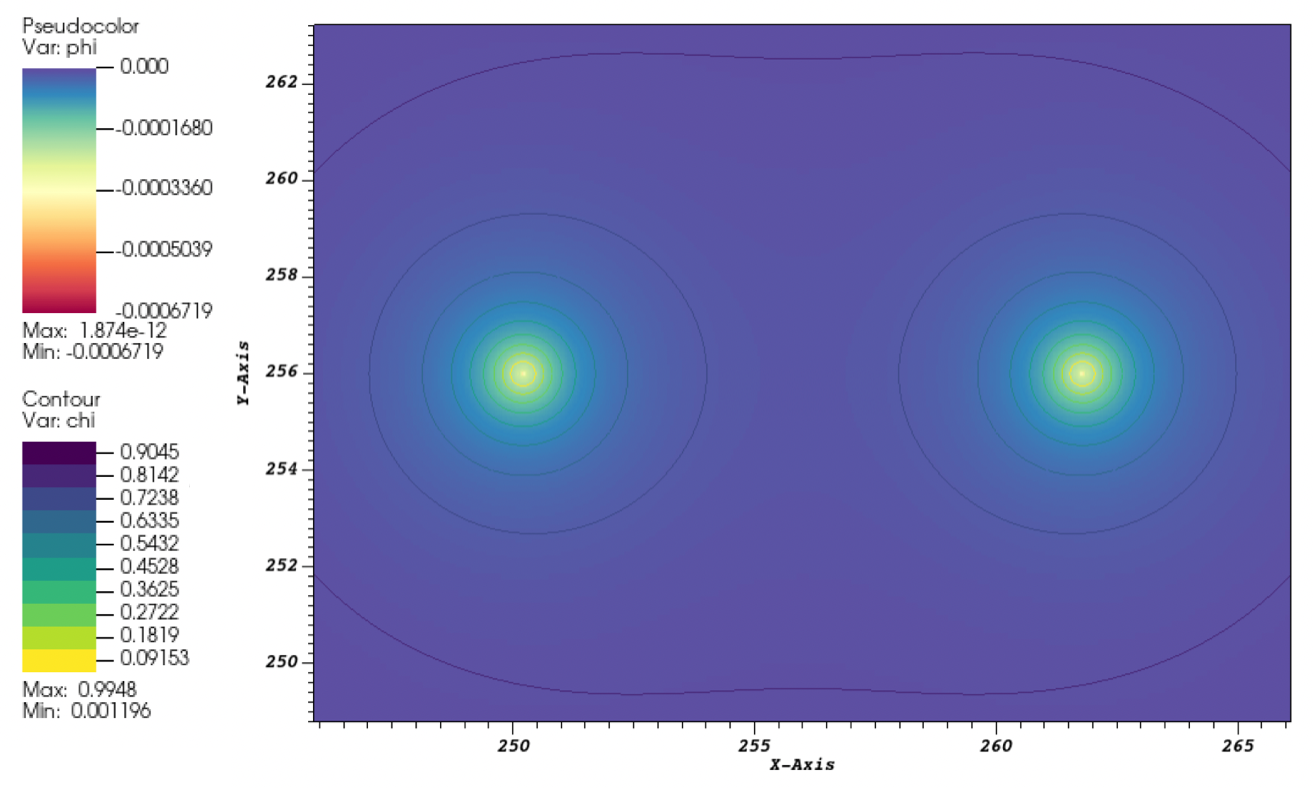}
    \end{center}
    \caption{Early-time distribution of the scalar conjugate momentum for an equal-charge binary with $M_1=M_2=0.5$, $q_1=q_2=0.1$ and $\alpha_0=3000$. The initial profile is given by Eq.~\eqref{config}, with $(A_0,r_0,\omega)=(0.0001,0.3,1.0)$.}
    \label{ID_a3000_q01}
\end{figure}

The scalar sector is initialized with
\begin{equation}
\phi(t=0)=0,
\end{equation}
while the conjugate momentum is prescribed as a small Gaussian seed,
\begin{equation}
\Pi(r)=\frac{A_0}{4\pi}\exp\left[-\left(\frac{r-r_0}{\omega}\right)^2\right] .
\label{config}
\end{equation}
Thus, the black holes are not initially scalarized, and any scalar hair that develops during the evolution is generated dynamically. Since $D_i\phi=0$ initially, the scalar sector does not source the momentum constraint, while its contribution to the Hamiltonian constraint is controlled by the small amplitude $A_0$ through the $\Pi^2$ term. The corresponding early-time distribution of the conjugate scalar momentum is
shown in Fig.~\ref{ID_a3000_q01} for the equal-charge configuration
\(q_1=q_2=0.1\) with coupling \(\alpha_0=3000\). The profile is localized around
the individual black holes and acts only as a small dynamical seed for the scalar
sector. Since the scalar field itself vanishes initially, the configuration does
not represent an initially scalarized binary; any nontrivial scalar hair observed
at later times is therefore generated dynamically by the electromagnetic--scalar
coupling.

The coupling function is specified by Eq.~\eqref{eq:coupling}. The simulations considered  in the main analysis (Sec.~IV.B) are summarized in Table~\ref{tab:simulations}. Cases I--IX isolate the effect of the coupling strength for a fixed equal-charge binary, whereas Cases VIII and X compare binaries with the same coupling but different total charge. The values of the coupling parameter \(\alpha_0\) were chosen using the scalarization thresholds and coupling dependence reported for isolated charged black holes in EMS theory~\cite{Fernandes:2019rez} as a guide. In the binary case, these values allow us to compare evolutions that end in descalarized and scalarized remnants while also testing the role of the charge retained after
coalescence.

\begin{table}[H]
\centering
\begin{tabular}{c c c c c}
\hline\hline
Case & $q_1$ & $q_2$ & $\alpha_0$ & Evolution outcome\\
\hline
I   & $+0.1$ & $+0.1$ & $600$  & descalarized \\
II  & $+0.1$ & $+0.1$ & $1000$ &  descalarized \\
III & $+0.1$ & $+0.1$ & $1500$ &  descalarized \\
IV   & $+0.1$ & $+0.1$ & $1800$  & descalarized \\
V & $+0.1$ & $+0.1$ & $1900$ &  descalarized \\
VI  & $+0.1$ & $+0.1$ & $2000$ &  scalarized \\
VII & $+0.1$ & $+0.1$ & $2500$ &  scalarized \\
VIII   & $+0.1$ & $+0.1$ & $3000$  & scalarized \\
IX  & $+0.1$ & $+0.1$ & $4000$ &  scalarized \\
X  & $+0.1$ & $-0.1$ & $3000$ &  descalarized \\
\hline\hline
\end{tabular}
\caption{Summary of the binary configurations evolved in this work. The quantities $q_1$ and $q_2$ denote the electric charges of the individual black holes, and $\alpha_0$ is the electromagnetic--scalar coupling parameter. The initial scalar seed is specified through the Gaussian profile imposed on the scalar conjugate momentum $\Pi$, as given in Eq.~\eqref{config}. The final state is classified according to the late-time behavior of the horizon-averaged scalar field on the remnant apparent horizon.}
\label{tab:simulations}
\end{table}

\subsection{Evolution setup}

All evolutions use a Courant factor $C=0.25$ and a Kreiss--Oliger dissipation parameter $\sigma=1$. The computational domain has an outer boundary at $L=512$,  where radiative (Sommerfeld-type) boundary conditions are imposed on the evolved variables to minimize spurious reflections. Reflection symmetry is imposed across the $z=0$ plane. The coarsest refinement level has a resolution of $dx=4$, and eight additional refinement levels are used with a refinement factor of two, giving an effective resolution of $dx\simeq0.0156$ near the apparent horizons. The moving-puncture gauge parameters are $\mu=3/4$ and $\eta=1$, and the electromagnetic damping parameters are $\lambda_1=\lambda_2=1$. Throughout this work, we use geometrized units ($G=c=1$). The quantity $M$
denotes the total ADM mass of the initial spatial hypersurface. All times and
lengths are normalized by $M$, which is set to unity in our simulations.

\section{Numerical results}
\label{results}

\subsection{Initial-data convergence and constraint monitoring}

\begin{figure}[H]
\centering
\includegraphics[width=0.9\linewidth]{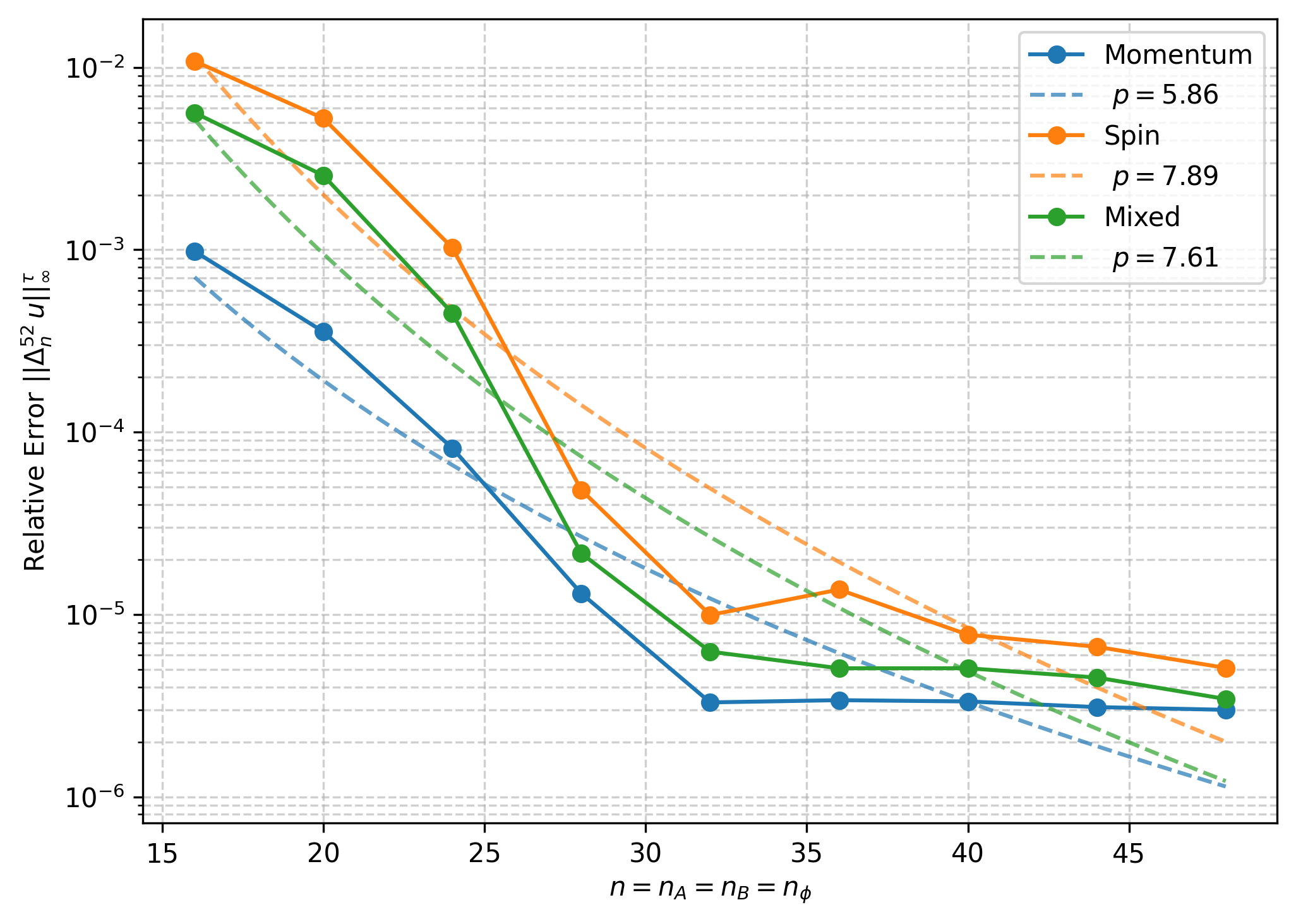}
\caption{Convergence test of the charged-puncture initial-data solver implemented in this work. The solver is based on the TwoPunctures method \cite{Ansorg:2004ds}, with extensions to charged black holes following Ref. \cite{Bozzola:2019aaw}. The error is computed with respect to a reference solution with $N=52$ using Eq.~\eqref{delta}.}
\label{fig:000}
\end{figure}

We first verify the convergence of the charged-puncture initial-data solver. Following Ref.~\cite{Bozzola:2019aaw}, the regular correction $u$ is computed using different numbers of Chebyshev collocation points and compared with a high-resolution solution. The relative infinite norm is
\begin{equation}
\norm{\Delta^N_n u}_{\infty}=\max\left|\frac{u^n(x)-u^N(x)}{u^N(x)}\right| .
\label{delta}
\end{equation}
For the test cases shown in Table~\ref{tab:BH_parameters}, the convergence behavior is shown in Fig.~\ref{fig:000}. The mixed case gives a slope close to $4.4$, indicating approximately fourth-order convergence for the initial-data solver in the parameter range considered.
\begin{table}[H]
\centering
\resizebox{\columnwidth}{!}{
\begin{tabular}{lcccccc}
\hline\hline
& \multicolumn{2}{c}{Momentum} & \multicolumn{2}{c}{Spin} & \multicolumn{2}{c}{Mixed} \\
Parameter & $BH_1$ & $BH_2$ & $BH_1$ & $BH_2$ & $BH_1$ & $BH_2$ \\
\hline
Mass $M_i$ & $0.5$ & $0.5$ & $0.5$ & $0.5$ & $0.5$ & $0.5$ \\
Charge $Q_i$ & $0.1$ & $0.1$ & $0.1$ & $0.1$ & $0.1$ & $0.1$ \\
Linear momentum $\vec P_i$ & $(0.1,0,0)$ & $(0,0.1,0)$ & $(0,0,0)$ & $(0,0,0)$ & $(0.1,0,0)$ & $(0,0.1,0)$ \\
Spin $\vec S_i$ & $(0,0,0)$ & $(0,0,0)$ & $(0,0,0.1)$ & $(0,0,-0.1)$ & $(0,0,0.1)$ & $(0,0,-0.1)$ \\
\hline\hline
\end{tabular}}
\caption{Initial parameters used for the convergence test of the charged-puncture solver. The binary separation is $D=8$ in all cases.}
\label{tab:BH_parameters}
\end{table}

\begin{figure}[H]
    \begin{center}
        \includegraphics[width=85mm]{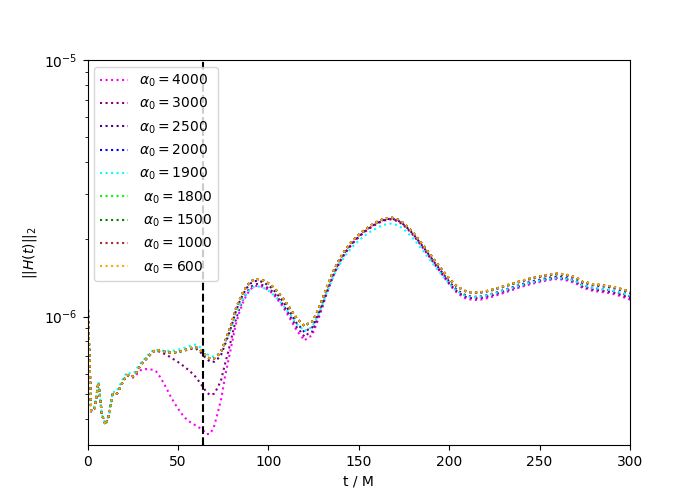}
        \includegraphics[width=85mm]{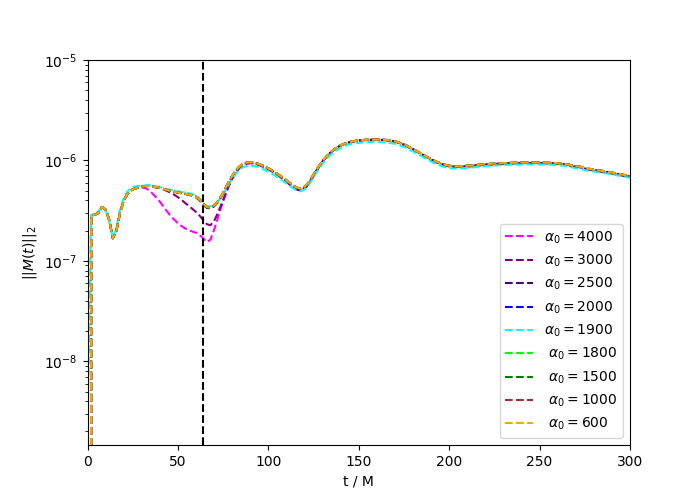}
        \includegraphics[width=85mm]{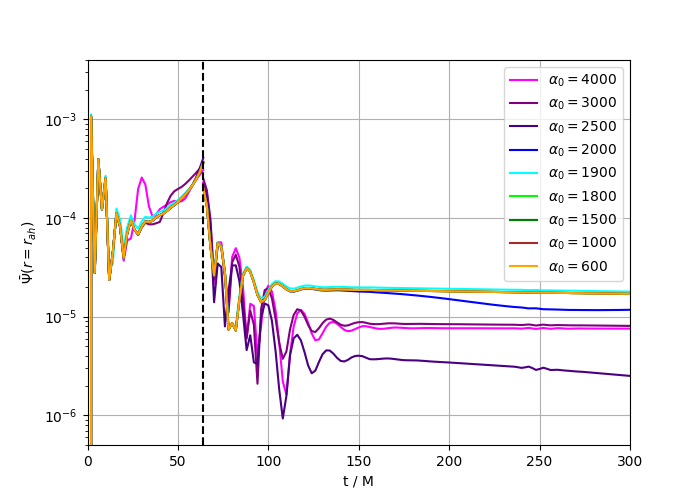}
    \end{center}
    \caption{Constraint monitoring for equal-charge binaries with $q_1=q_2=0.1$, comparing different values of $\alpha_0$. Top panel: $L_2$ norms of the Hamiltonian constraint violation. Middle panel: $L_2$ norms of the momentum constraint violation. Bottom panel: horizon-averaged value of the electromagnetic constraint-damping variable $\Psi$ measured on the apparent horizons. The vertical dashed line indicates the formation of the common apparent horizon.}
    \label{rest}
\end{figure}

For time evolutions, Fig.~\ref{rest} shows the $L_2$ norms of the Hamiltonian and the momentum constraints for equal-charge binaries with $q_1=q_2=0.1$, comparing different values of $\alpha_0$. The same figure also shows the horizon-averaged value of the electromagnetic constraint-damping variable $\Psi$. The constraints remain under control throughout the simulations, including during the merger. These diagnostics support the numerical stability of the evolutions; a systematic resolution study of the full nonlinear EMS evolutions would be required to determine the convergence order of the dynamical simulations themselves.

\subsection{Dynamical scalarization and descalarization}

To characterize the scalar field dynamics, we monitor the horizon-averaged scalar field
\begin{equation}
\bar{\phi}_{\mathcal H_A}= \frac{1}{A}\int_{\mathcal H_A}|\phi|\,dA,
\label{eq:phiavg}
\end{equation} 
Before the merger, this quantity is evaluated on the individual horizons, while after the formation of a common horizon, it is evaluated on the remnant.

\begin{figure}[H]
    \begin{center}
        \includegraphics[width=95mm]{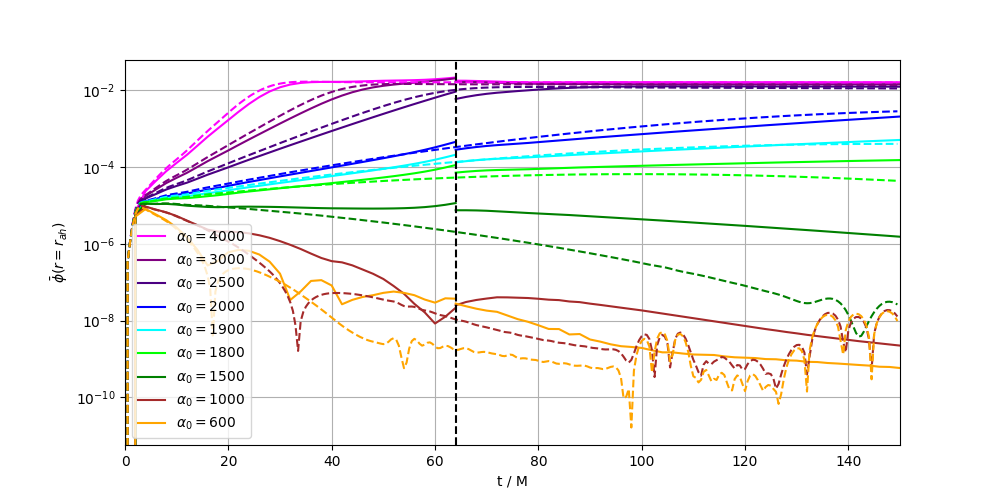}
    \end{center}
    \caption{Scalar field averaged over the apparent horizons during the merger of equal-charge black holes with $q_1=q_2=0.1$ and initial separation $d = 12M$. The vertical dashed line indicates the formation of the common apparent horizon. Before this time, the scalar field is extracted from one of the individual black-hole horizons; afterwards, it is extracted from the remnant horizon. For $\alpha_0=600, 1000, 1500, 1800$, and $1900$, the remnant descalarizes, whereas for $\alpha_0=2000, 2500, 3000$, and $4000$, it retains a nonzero scalar field. Solid curves correspond to the binary evolutions,
while dashed curves correspond to isolated charged black holes with masses and
charges similar to those of the post-merger remnant.}
    \label{scalarization}
\end{figure}

Fig.~\ref{scalarization} shows {\bf{$\bar{\phi}_{\mathcal H_A}$}} 
for equal-charge binaries with $q_1=q_2=0.1$.  For $\alpha_0=600, 1000, 1500, 1800$, and $1900$, the scalar field grows during the early stages of the evolution. For the cases $\alpha_0=600, 1000$, and $1500$, the subsequent decrease is already apparent within the time interval shown in the figure, indicating that the remnant dynamically descalarizes. For $\alpha_0=1800$ and $1900$, the scalar field continues to increase slowly until the end of the displayed time interval. At later times, however, it reaches a maximum and subsequently decreases, indicating that the remnant ultimately dynamically descalarizes. We therefore classify the cases $\alpha_0=1800$ and $1900$ as descalarized based on their later-time evolution, although this behavior is not fully visible in the time window shown in Fig.~\ref{scalarization}. In contrast, for $\alpha_0=2000, 2500, 3000$, and $4000$, the scalar field remains finite on the common horizon and approaches an approximately constant late-time value, indicating a scalarized remnant (solid curves). The results suggest the existence of a transition  region between  $\alpha_0=1900$ and $\alpha_0=2000$ that separates scalarized and descalarized merger remnants. Thus, the values of $\alpha_0$ considered here lie on different sides of a transition region separating descalarized and scalarized post-merger outcomes. This behavior is qualitatively consistent with dynamical scalarization and descalarization in isolated EMS black holes~\cite{Xiong:2022ozw,Niu:2022zlf,Jiang:2023yyn}, but here the final state is determined by the properties of the merger remnant rather than by a single initial black hole. Additionally, to compare the binary transition with the isolated-black-hole expectation, we
also evolved isolated charged black holes with masses and charges similar to those of the
post-merger remnant (dashed curves). The isolated evolutions show a transition in the same range
of coupling values, \(1900< \alpha_0 <2000\), as the binary
remnants. This agreement indicates that the binary outcome is primarily
controlled by whether the remnant parameters allow the final black hole to
support the scalarized branch. Small differences between the isolated and binary
evolutions are expected because the merger remnant is initially dynamical and
nonstationary.

\begin{figure}[H]
    \begin{center}
        \includegraphics[width=88mm]{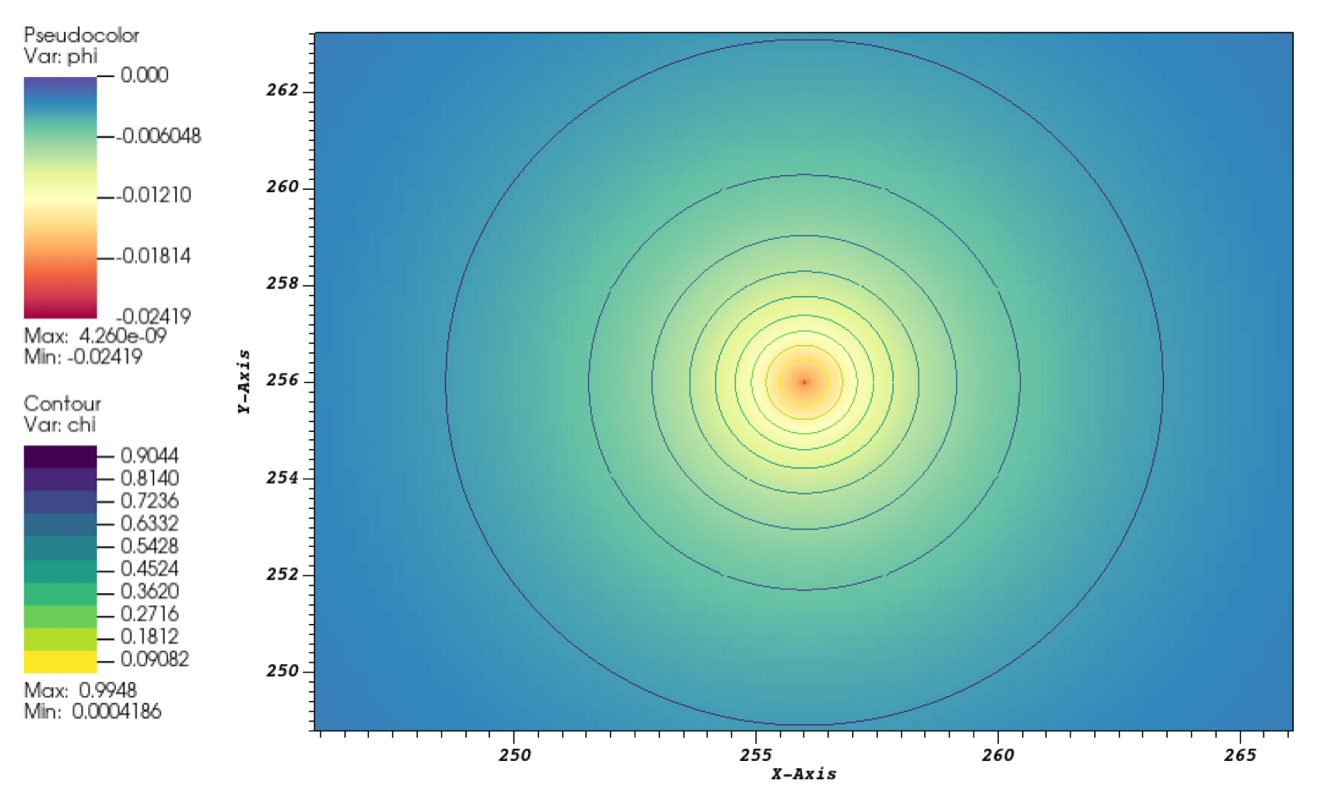}
    \end{center}
    \caption{Distribution of the scalar field shortly after merger at $t=80$ for an equal-charge binary with $\alpha_0=3000$. The remnant black hole is scalarized.}
    \label{t=80_a3000_q01}
\end{figure}

Fig.~\ref{t=80_a3000_q01} shows the scalar field shortly after the merger for $\alpha_0=3000$ and $q_1=q_2=0.1$. The scalar field remains localized around the common horizon and decreases toward spatial infinity, consistent with a scalarized remnant.

\begin{figure}[H]
    \begin{center}
        \includegraphics[width=95mm]{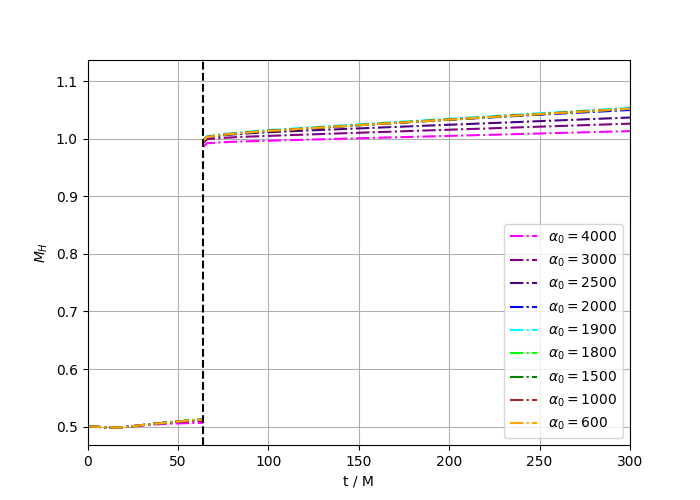}
    \end{center}
    \caption{Time evolution of the apparent-horizon masses during the binary coalescence for different values of $\alpha_0$. The vertical dashed line indicates the formation of the common apparent horizon. To the left of this line, the horizon mass is extracted from one of the individual black holes, while to the right it is extracted from the remnant black hole.}
    \label{mass_SI}
\end{figure}

The horizon masses in Fig.~\ref{mass_SI} show the expected behavior of a head-on coalescence. Before the merger, the individual apparent-horizon masses remain approximately constant. After the common apparent horizon forms, the remnant mass approaches $M_{\rm rem}\simeq1$, up to the energy carried away by gravitational, electromagnetic, and scalar radiation. Since the initial data have vanishing spin and vanishing orbital angular momentum, the relevant remnant quantities for the scalarization process are $M_{\rm rem}$, $Q_{\rm rem}$, and the effective charge-to-mass ratio.

In EMS theory, the conserved electric charge is associated with the flux of $f(\phi) ^\star F$. We define the horizon charge as
\begin{equation}
\tilde{Q}_H = \frac{1}{4 \pi} \oint_{\mathcal{H}} f(\phi) ^\star F \,,
\end{equation}
where the integral is evaluated over the apparent horizon $\mathcal{H}$.

\begin{figure}[H]
    \begin{center}
        \includegraphics[width=95mm]{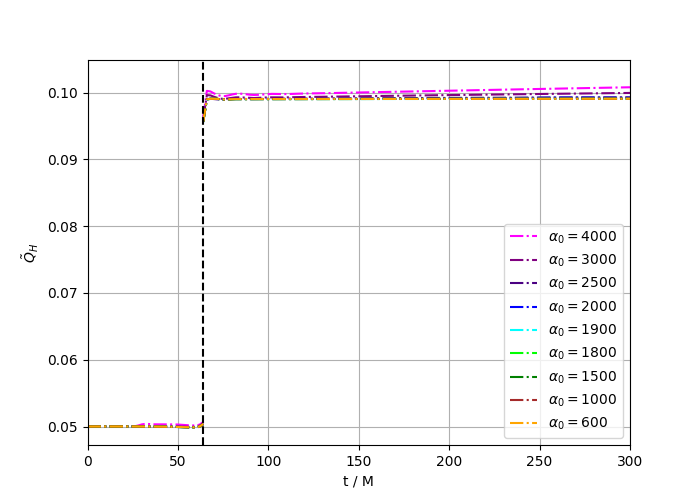}
    \end{center}
    \caption{Evolution of the horizon charge $\tilde{Q}_H$ during the binary coalescence for different values of the coupling parameter $\alpha_0$. The vertical dashed line marks the formation of the common apparent horizon. To the left of this line, the charge is measured on one of the individual black-hole horizons, while to the right it is measured on the remnant horizon. For all couplings the remnant charge remains close to the initial total charge.}
    \label{charge}
\end{figure}

Figure~\ref{charge} shows the evolution of the horizon charge for different values of the coupling parameter $\alpha_0$. Before the merger, the charge is computed on one of the individual black-hole horizons, while after the formation of the common apparent horizon, it is evaluated on the remnant horizon.  For all couplings, the remnant horizon charge remains close to the initial total charge of the binary, $\tilde{Q}_H \simeq 0.1$. A small increase in the horizon charge is observed after the merger, with its magnitude depending on $\alpha_0$, consistent with the nontrivial electromagnetic--scalar coupling in the scalarized remnant.

\subsection{Equal-charge versus opposite-charge binaries}

\begin{figure}[H]
    \begin{center}
        \includegraphics[width=95mm]{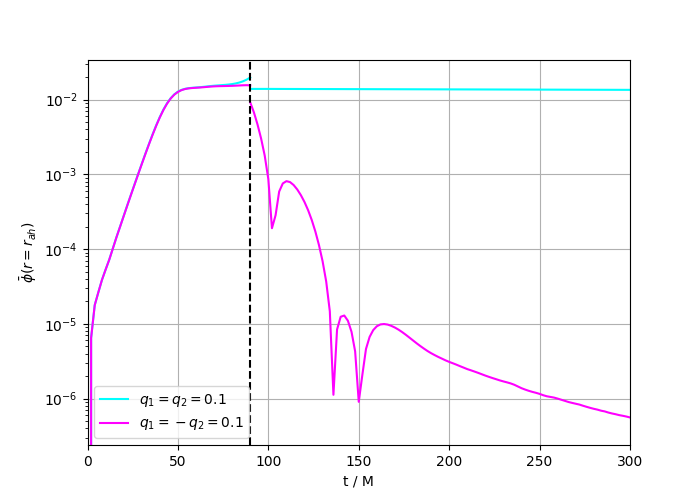}
    \end{center}
    \caption{Evolution of scalarization during the merger of charged black holes for $\alpha_0=3000$. The individual black holes scalarize before coalescence, but the scalarization of the remnant depends on its final charge. An initial separation of $d = 20M$ is considered. For $q_1=q_2=0.1$ the remnant remains scalarized, whereas for $q_1=-q_2=0.1$ the scalar hair is radiated away and the final black hole is effectively bald.}
    \label{scl}
\end{figure}

Fig.~\ref{scl} compares the scalar evolution for $\alpha_0=3000$ in equal-charge and opposite-charge binaries. In both cases, the individual black holes scalarize before merger. 
The initial separation is set to \(d=20M\), allowing the scalarization of each
black hole to develop before the collision.
However, their post-merger behavior is different. For $q_1=q_2=0.1$, the remnant retains a nonzero net electric charge and remains scalarized. For $q_1=-q_2=0.1$, charge cancellation strongly suppresses the net charge of the remnant, weakening the scalar source term. The scalar field is then radiated away or absorbed by the final horizon, and the remnant dynamically relaxes toward an effectively bald configuration. This charge-cancellation effect is also the key distinction between same-sign and opposite-sign charged black-hole collisions in pure Einstein--Maxwell theory~\cite{Zilhao:2012gp,Zilhao:2013nda}.

This comparison provides the main physical interpretation of the simulations. Pre-merger scalarization of the individual horizons does not guarantee a scalarized final state. The outcome is controlled primarily by the ability of the remnant to support scalar hair, which in EMS theory is tied to the electromagnetic invariant and the retained charge. This conclusion is consistent with dynamical descalarization studies of isolated EMS black holes, where the effective charge-to-mass ratio controls whether scalar hair can be maintained~\cite{Niu:2022zlf}. It also parallels the scalarized/descalarized remnant phenomenology observed in scalar-Gauss--Bonnet BBH mergers in the decoupling limit~\cite{Silva:2020omi,Elley:2022ept}, although the source of the scalar field is different: electromagnetic in EMS theory and curvature-driven in scalar-Gauss--Bonnet gravity.

\subsection{Radiation and scalarized remnant}

\begin{figure}[H]
    \begin{center}
        \includegraphics[width=95mm]{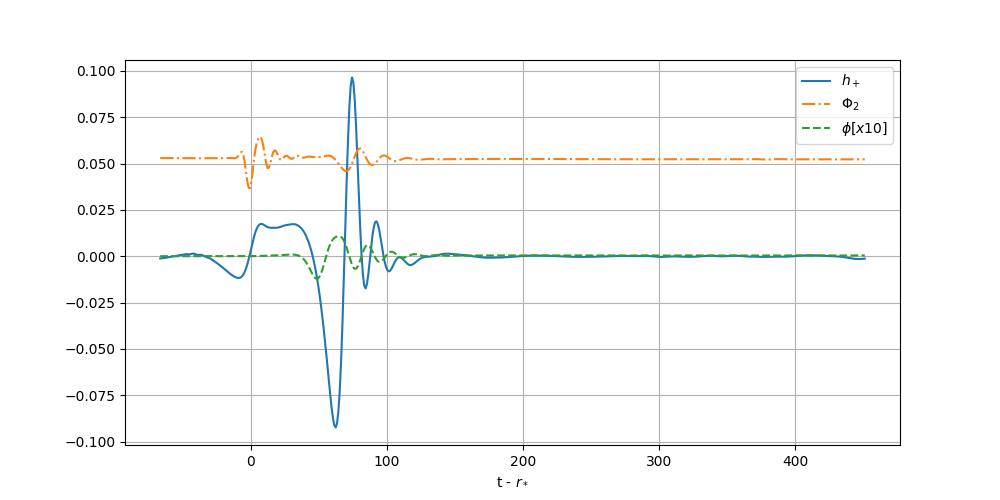}
    \end{center}
    \caption{Gravitational, electromagnetic and scalar radiation produced during the head-on collision for equal-charge $(q_1=q_2=0.1)$ binaries and $\alpha_0=3000$. An initial separation of $d=12M$ is considered. The gravitational waveform is extracted from the Newman--Penrose scalar $\Psi_4$, while the electromagnetic and scalar signals are obtained from the corresponding evolved fields at the extraction radius.}
    \label{str}
\end{figure}

Fig.~\ref{str} displays the dominant gravitational-wave mode, extracted from the Newman--Penrose scalar $\Psi_4$ and decomposed into spherical harmonics, together with representative electromagnetic and scalar signals for $\alpha_0=3000$ and $q_1=q_2=0.1$. The first $\sim40$ time units are dominated by junk radiation associated with relaxation of the initial data. After this transient, the scalar and gravitational signals develop correlated features during the nonlinear stage of the collision, especially near the merger. The simultaneous presence of gravitational, electromagnetic, and scalar channels distinguishes EMS evolutions from pure Einstein--Maxwell charged-black-hole collisions: in EMS theory, the electromagnetic sector not only radiates but also sources scalar radiation through the nonminimal coupling~\cite{Zilhao:2012gp}.

\begin{figure}[H]
    \begin{center}
        \includegraphics[width=95mm]{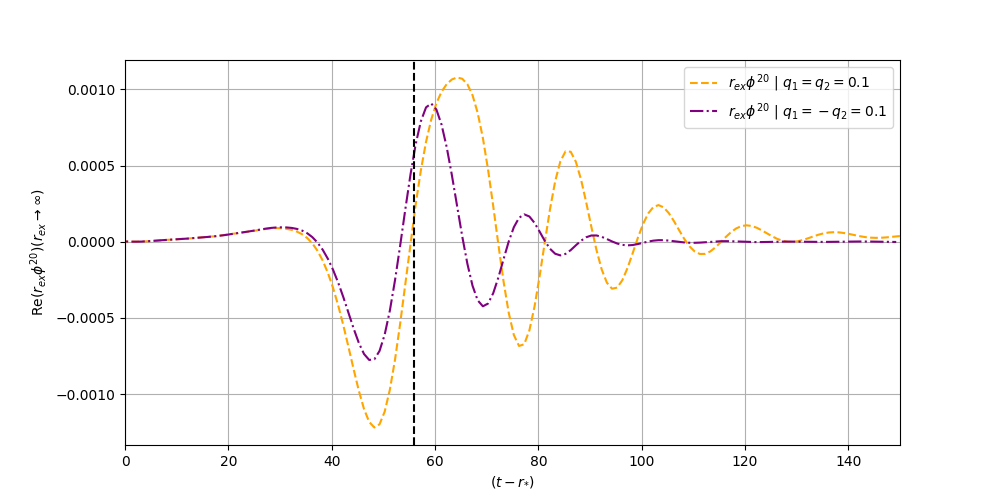}
    \end{center}
    \caption{Evolution of the $(\ell=2,m=0)$ scalar-field mode $\phi$, extrapolated to infinite extraction radius, for equal-charge $(q_1=q_2=0.1)$ and opposite-charge $(q_1=-q_2=0.1)$ binaries and $\alpha_0=3000$. An initial separation of $d=12M$ is considered. The vertical dashed line marks the time at which the common apparent horizon forms, signaling the merger of the two black holes.}
    \label{strain}
\end{figure}

The asymptotic scalar radiation for equal-charge $(q_1=q_2=0.1)$ and opposite-charge $(q_1=-q_2=0.1)$ binaries and $\alpha_0=3000$ is shown in Fig. \ref{strain}, where the $(\ell=2,m=0)$ mode of the scalar field is extrapolated to an infinite extraction radius. The vertical dashed line marks the formation of the common apparent horizon. Unlike the gravitational and electromagnetic signals, no significant burst of spurious initial scalar radiation is observed. This is expected because the scalar field is absent in the initial data, and its growth is driven by the physical tachyonic instability associated with the EMS coupling, rather than by a numerical relaxation of the initial configuration. This provides further evidence that the individual black holes dynamically scalarize prior to the merger.

After the formation of the common horizon, the equal-charge binary $(q_1=q_2=0.1)$ exhibits a more pronounced oscillatory signal. This behavior is consistent with the relaxation of a scalarized remnant black hole, whose scalar-field configuration continues to adjust while approaching its final equilibrium state. In contrast, for the opposite-charge binary $(q_1=-q_2=0.1)$, the scalar signal decays rapidly after merger. In this case, charge cancellation suppresses the electromagnetic source responsible for sustaining scalar hair, leading to the dynamical descalarization of the remnant.

Therefore, at least for the configurations considered here, the extracted scalar radiation provides a clear diagnostic of the post-merger state, allowing one to distinguish between remnants that remain scalarized and those that dynamically descalarize after the merger.

In contrast to BBH evolutions embedded in an external scalar-field environment~\cite{Healy:2011ef}, where scalar radiation is sourced by the ambient scalar field, the scalar radiation observed here is generated intrinsically by the EMS dynamics through the electromagnetic--scalar coupling.

\begin{figure}[H]
    \begin{center}
        \includegraphics[width=85mm]{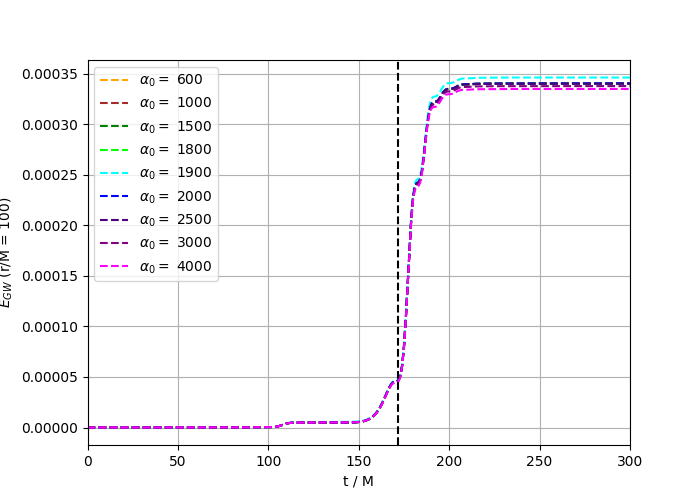}
        \includegraphics[width=86mm]{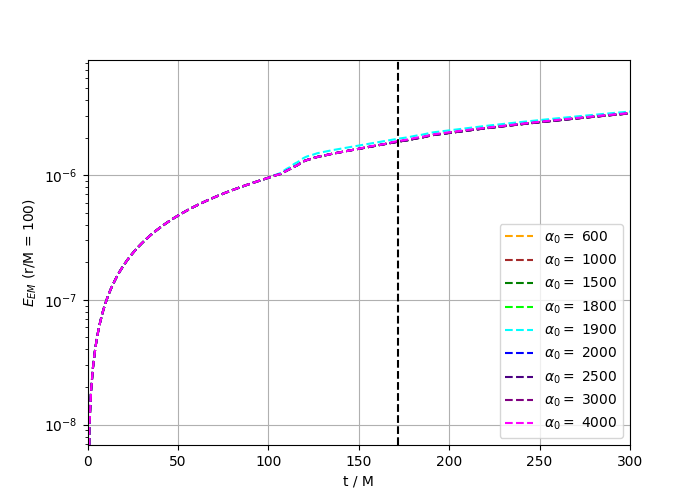}
        \includegraphics[width=77mm]{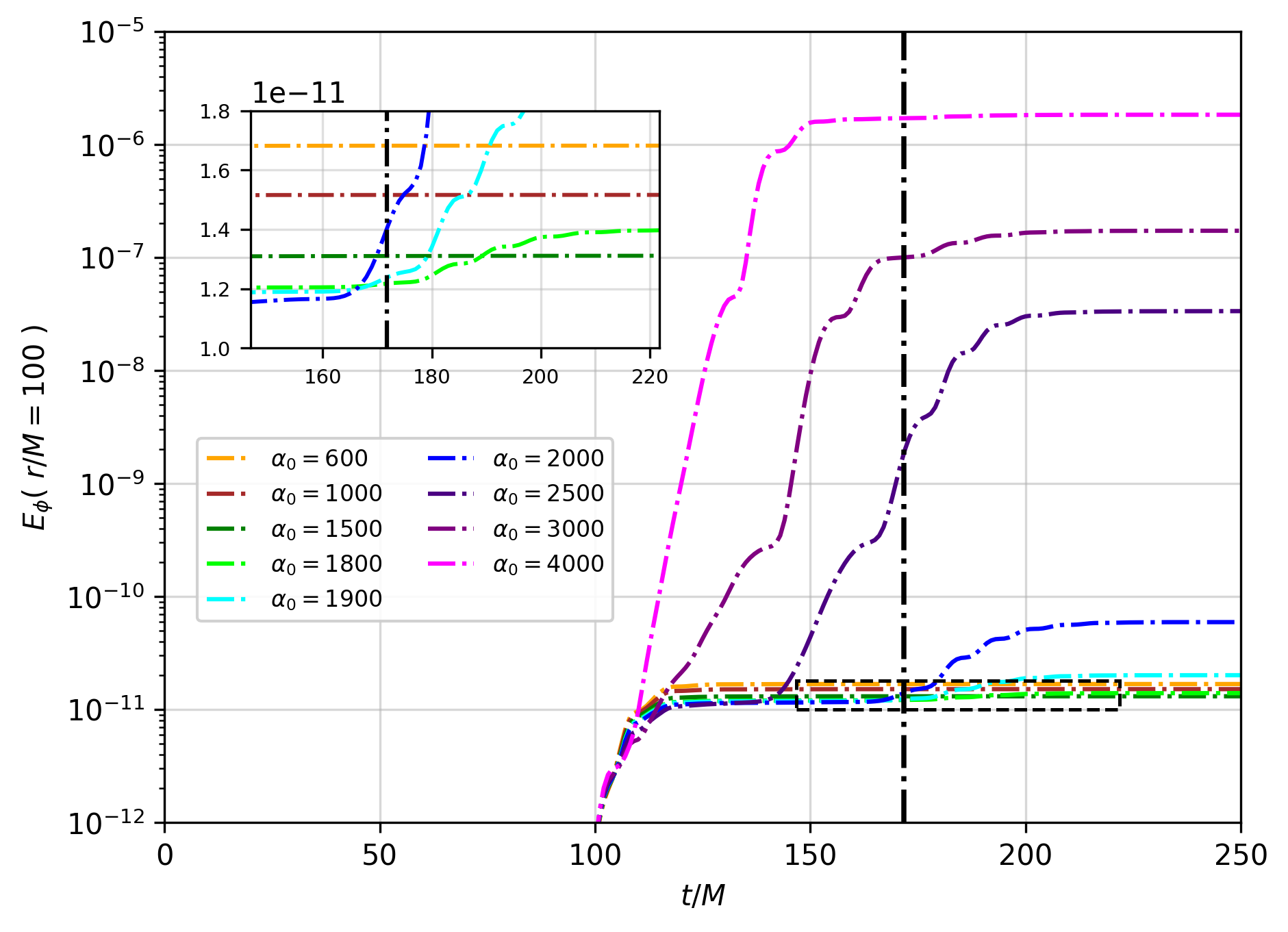}
    \end{center}
    \caption{Cumulative radiated energies extracted at \(r/M=100\) for the equal-charge
binary configurations with \(q_1=q_2=0.1\) and different values of the coupling
parameter \(\alpha_0\). Top panel: gravitational radiated energy. Middle panel:
electromagnetic radiated energy. Bottom panel: scalar radiated energy. The vertical
dashed line marks the approximate arrival time of the merger signal at the
extraction radius. The gravitational and electromagnetic channels show only a
weak dependence on \(\alpha_0\), whereas the scalar channel is strongly enhanced
for the scalarized remnants and remains close to the initial-seed level for the
descalarized cases.}
    \label{radiatedenergies}
\end{figure}

To further quantify the role of the different radiative channels, we also
computed the cumulative radiated energies in the gravitational, electromagnetic,
and scalar sectors at the extraction radius \(r/M=100\). The corresponding
results are shown in Fig.~\ref{radiatedenergies}. The gravitational and
electromagnetic radiated energies display only a weak dependence on the coupling
parameter \(\alpha_0\) for the configurations considered here. In contrast, the
scalar radiated energy shows a clear dependence on the scalarization state of
the remnant. For \(\alpha_0\leq 1900\), the scalar energy remains close to the
level associated with the initial scalar seed, whereas for larger couplings it
increases significantly, consistently with the formation of a scalarized
remnant. The case $\alpha_0=2000$ lies close to the transition region and
therefore display an intermediate behavior. This confirms that the scalar channel is the radiative sector most directly
sensitive to the scalarization/descalarization transition. We stress, however,
that in the present work the radiated energies are used as a diagnostic of the
post-merger scalar state, rather than as a complete energy-budget analysis of
the coalescence.

\begin{figure}[H]
\centering
\includegraphics[width=1.00\linewidth]{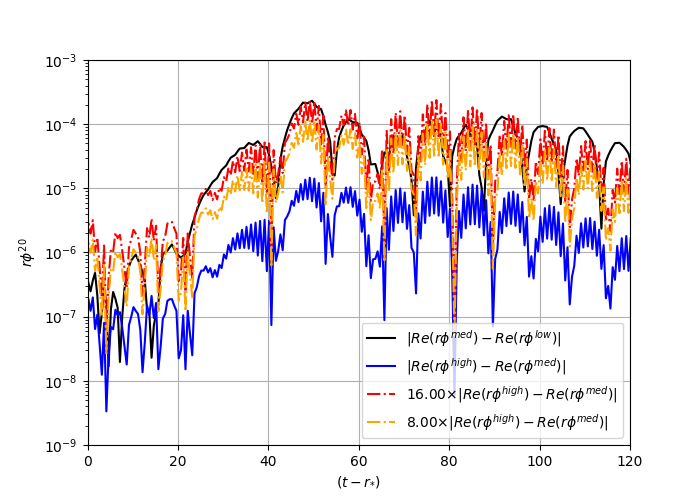}
\caption{Convergence test of head-on collision.
Differences in the real part of the
$\ell=2, m = 0$ mode of $r\phi^{20}$ between three different resolutions. 
We also show the data rescaled
by a factor consistent with either third $(\times8)$ or fourth $(\times16)$ order convergence, following Eq.~\eqref{Q}.}
\label{fig:Convergence}
\end{figure}

The evolution of the head-on collision is additionally validated using convergence test for low, medium and high resolution, $\Delta x_{low} = 16M$, $\Delta x_{med} = 8M$ and $\Delta x_{high} = 4M$, respectively, in the coarsest resolution level, Fig \ref{fig:Convergence}. The order convergence $Q$ is
\begin{equation}
\frac{|r\phi^{low} - r\phi^{med}|}{|r\phi^{med} - r\phi^{high}|} = \frac{|\Delta x_{low}^{Q} - \Delta x_{med}^{Q}|}{|\Delta x_{med}^{Q} - \Delta x_{high}^{Q}|} = Q.
\label{Q}
\end{equation}

\section{Conclusions}
\label{Conclusions}

We have performed numerical-relativity simulations of head-on binary black hole coalescences in Einstein--Maxwell--Scalar theory. The binaries were constructed from charged puncture initial data representing two Reissner--Nordstr\"om black holes immersed in a purely kinetic scalar perturbation, with the scalar field initially vanishing and its conjugate momentum providing the initial seed. We then evolved the coupled gravitational, electromagnetic, and scalar sectors and monitored the apparent horizons, the scalar field on the horizons, and the emitted radiation. The main purpose of the analysis was to determine whether the electromagnetic--scalar coupling can dynamically generate scalar hair in a binary spacetime, and whether such hair survives or is lost after the merger.

Our simulations show that the nonminimal electromagnetic--scalar coupling can dynamically source a nontrivial scalar configuration even when the individual black holes are initially scalar-free. For the quadratic coupling considered here, the purely kinetic seed can trigger tachyonic growth of the scalar field near the horizons, in agreement with the scalarization mechanism expected for charged black holes in Class IIA EMS models.  The post-merger fate of the scalar field depends on the coupling parameter $\alpha_0$ and on the charge-to-mass ratio of the remnant. For the equal charge-to-mass ratio configurations considered here, we identify a transition region $1900<\alpha_0<2000$: the remnant descalarizes for $\alpha_0\leq1900$, whereas it retains nonzero scalar hair for $\alpha_0\geq2000$. More generally, sufficiently small remnant charge-to-mass ratios prevent the scalarized configuration from being sustained, causing the scalar hair to decay through radiation and horizon absorption.

 A central result is the contrast between binaries with equal and opposite charge-to-mass ratios. For the equal-sign configurations, the remnant retains a nonzero electric charge and therefore a sufficiently strong electromagnetic source to support scalar hair.  By contrast, for opposite-sign configurations, charge cancellation during merger suppresses the scalar source in the final state. Although the individual black holes may scalarize before coalescence, the post-merger remnant cannot sustain the scalar field and dynamically relaxes toward an effectively bald black hole. Thus, the decisive factor is not only the presence of scalar hair before the merger, but whether the remnant itself lies in the scalarized sector of the theory.

These results extend previous studies of dynamical scalarization and descalarization in isolated EMS black holes~\cite{Xiong:2022ozw,Niu:2022zlf,Jiang:2023yyn} to a genuinely binary and strongly nonlinear merger spacetime. They are also qualitatively analogous to scalarized and descalarized remnants in scalar-Gauss--Bonnet BBH mergers~\cite{Silva:2020omi,Elley:2022ept}. The physical mechanism is nevertheless different: in EMS theory, the scalar field is sourced by the electromagnetic sector, whereas in curvature-coupled models, it is sourced by curvature invariants and, in some regimes, by spin.

The merger also excites scalar radiation whose time profile is qualitatively correlated with the dominant gravitational-wave mode during the nonlinear stage of the collision. This suggests that the scalar channel is dynamically sourced during coalescence rather than being only a remnant of the initial perturbation. From the numerical point of view, the simulations demonstrate that the GRChombo infrastructure can be adapted to evolve charged BBH spacetimes in EMS theory, with the constraint violations remaining under control in the evolutions considered here.

The scalar radiation extracted at infinity also provides a useful diagnostic of the post-merger state. For the configurations considered here, remnants that remain scalarized exhibit a long-lived oscillatory scalar signal, whereas dynamically descalarized remnants show a rapid decay of the scalar radiation. This suggests that scalar-wave observations could provide direct information about the scalarization state of the final black hole.

Several extensions are natural. A broader scan in $(q_1,q_2,\alpha_0)$ is needed to map the transition surface between scalarized and descalarized remnants and to compare it with the thresholds of isolated charged black holes. Quasi-circular binaries and rotating charged black holes should also be considered to determine the role of orbital angular momentum and spin. Finally, improved waveform extraction, radiated-flux calculations, and a systematic resolution study of the full nonlinear evolutions would allow a quantitative assessment of the scalar and electromagnetic imprints on the gravitational signal.\\

\acknowledgments

We thank the anonymous referee for valuable comments and suggestions. This work was partially supported by ANID Chile through FONDECYT Grant No 1220871  (P. A. G., Y. V., and J. D.). Y. V. acknowledges the financial support of DIDULS/ULS, through the project No PR25538511. The authors wish to thank the FIULS 2030 project 18ENI2-104235 -- CORFO for providing computing resources. Powered@NLHPC: This research was partially supported by the supercomputing infrastructure of the NLHPC (CCSS210001). P.A.G. would like to thank the Facultad de Ciencias, Universidad de La Serena for its hospitality.


\begin{thebibliography}{99}


\bibitem{LIGOScientific:2016aoc}
B.~P.~Abbott \textit{et al.} [LIGO Scientific and Virgo],
``Observation of Gravitational Waves from a Binary Black Hole Merger,''
Phys. Rev. Lett. \textbf{116} (2016) no.6, 061102
[arXiv:1602.03837 [gr-qc]].




\bibitem{Araya:2022few}
I.~J.~Araya, N.~D.~Padilla, M.~E.~Rubio, J.~Sureda, J.~Maga{\~n}a and L.~Osorio,
``Dark matter from primordial black holes would hold charge,''
JCAP \textbf{02} (2023), 030
[arXiv:2207.05829 [astro-ph.CO]].






\bibitem{Moreno:2021neu}
C.~Moreno, J.~C.~Degollado, D.~N{\'u}{\~n}ez and C.~Rodr{\'\i}guez-Leal,
``Gravitational and Electromagnetic Perturbations of a Charged Black Hole in a General Gauge Condition,''
Particles \textbf{4} (2021) no.2, 106-128
[arXiv:2104.11742 [gr-qc]].




\bibitem{Palenzuela:2009hx}
C.~Palenzuela, L.~Lehner and S.~Yoshida,
``Understanding possible electromagnetic counterparts to loud gravitational wave events: Binary black hole effects on electromagnetic fields,''
Phys. Rev. D \textbf{81} (2010), 084007
[arXiv:0911.3889 [gr-qc]].



\bibitem{Astefanesei:2019pfq}
D.~Astefanesei, C.~Herdeiro, A.~Pombo and E.~Radu,
``Einstein-Maxwell-scalar black holes: classes of solutions, dyons and extremality,''
JHEP \textbf{10} (2019), 078
[arXiv:1905.08304 [hep-th]].


\bibitem{Hirschmann:2017psw}
E.~W.~Hirschmann, L.~Lehner, S.~L.~Liebling and C.~Palenzuela,
``Black Hole Dynamics in Einstein-Maxwell-Dilaton Theory,''
Phys. Rev. D \textbf{97} (2018) no.6, 064032
[arXiv:1706.09875 [gr-qc]].





\bibitem{Flanagan:1997sx}
E.~E.~Flanagan and S.~A.~Hughes,
``Measuring gravitational waves from binary black hole coalescences: 1. Signal-to-noise for inspiral, merger, and ringdown,''
Phys. Rev. D \textbf{57} (1998), 4535-4565
[arXiv:gr-qc/9701039 [gr-qc]].






\bibitem{Benjamin}
Benjamin Aylott et al 2009, "Testing gravitational-wave searches with numerical relativity waveforms: results from the first Numerical INJection Analysis (NINJA) project” Class. Quantum Grav. 26 165008.


\bibitem{Blanchet:2013haa}
L.~Blanchet,
``Post-Newtonian Theory for Gravitational Waves,''
Living Rev. Rel. \textbf{17} (2014), 2
[arXiv:1310.1528 [gr-qc]].






\bibitem{Alcubierre}
Miguel Alcubierre. Introduction to 3+1 Numerical Relativity. Oxford University Press, Apr. 2008. isbn:
9780199205677. 

\bibitem{Gourgoulhon:2007ue}
E.~Gourgoulhon,
``3+1 formalism and bases of numerical relativity,''
[arXiv:gr-qc/0703035 [gr-qc]].



 
\bibitem{Baumgarte}
"Thomas W. Baumgarte and Stuart L. Shapiro. “Solving the Constraint Equations”. In: Numerical Relativity: Starting from Scratch", Cambridge University Press, 2021, pp. 87–105. 


\bibitem{Arnowitt:1962hi}
R.~L.~Arnowitt, S.~Deser and C.~W.~Misner,
``The Dynamics of general relativity,''
Gen. Rel. Grav. \textbf{40} (2008), 1997-2027
[arXiv:gr-qc/0405109 [gr-qc]].


\bibitem{Lindblom:2005qh}
L.~Lindblom, M.~A.~Scheel, L.~E.~Kidder, R.~Owen and O.~Rinne,
``A New generalized harmonic evolution system,''
Class. Quant. Grav. \textbf{23} (2006), S447-S462
[arXiv:gr-qc/0512093 [gr-qc]].



\bibitem{Alcubierre:2009ij}
M.~Alcubierre, J.~C.~Degollado and M.~Salgado,
``The Einstein-Maxwell system in 3+1 form and initial data for multiple charged black holes,''
Phys. Rev. D \textbf{80} (2009), 104022
[arXiv:0907.1151 [gr-qc]].


\bibitem{Pretorius:2005gq}
F.~Pretorius,
``Evolution of binary black hole spacetimes,''
Phys. Rev. Lett. \textbf{95} (2005), 121101
[arXiv:gr-qc/0507014 [gr-qc]].


\bibitem{Combi:2024inn}
L.~Combi and S.~M.~Ressler,
``Binary black hole metric approximation from inspiral to merger,''
Phys. Rev. D \textbf{113} (2026) no.4, 044023
[arXiv:2403.13308 [gr-qc]].


\bibitem{Aswathi:2025nxa}
P.~S.~Aswathi, W.~E.~East, N.~Siemonsen, L.~Sun and D.~Jones,
``Ultralight boson constraints from gravitational wave observations of spinning binary black holes,''
Phys. Rev. D \textbf{112} (2025) no.12, 123048
[arXiv:2507.20979 [gr-qc]].




\bibitem{Gliorio:2026yvh}
S.~Gliorio, M.~Della Rocca, S.~Barsanti, L.~Gualtieri, A.~Maselli and T.~P.~Sotiriou,
``Adiabatic evolution of asymmetric binaries on generic orbits with new fundamental fields I: characterization of gravitational wave fluxes,''
[arXiv:2603.10116 [gr-qc]].

\bibitem{Acevedo:2026xol}
J.~F.~Acevedo and A.~Ritz,
``Binary-boosted Dark Matter,''
[arXiv:2603.08781 [hep-ph]].


\bibitem{Mobilia:2026dxt}
L.~Mobilia, T.~Dal Canton and G.~M.~Guidi,
``Deep Learning Search for Gravitational Waves from Compact Binary Coalescence,''
[arXiv:2603.09386 [gr-qc]].








\bibitem{Healy:2011ef}
J.~Healy, T.~Bode, R.~Haas, E.~Pazos, P.~Laguna, D.~Shoemaker and N.~Yunes,
``Late Inspiral and Merger of Binary Black Holes in Scalar-Tensor Theories of Gravity,''
Class. Quant. Grav. \textbf{29} (2012), 232002
[arXiv:1112.3928 [gr-qc]].


\bibitem{Silva:2020omi}
H.~O.~Silva, H.~Witek, M.~Elley and N.~Yunes,
``Dynamical Descalarization in Binary Black Hole Mergers,''
Phys. Rev. Lett. \textbf{127} (2021) no.3, 031101
[arXiv:2012.10436 [gr-qc]].


\bibitem{Elley:2022ept}
M.~Elley, H.~O.~Silva, H.~Witek and N.~Yunes,
``Spin-induced dynamical scalarization, descalarization, and stealthness in scalar-Gauss-Bonnet gravity during a black hole coalescence,''
Phys. Rev. D \textbf{106} (2022) no.4, 044018
[arXiv:2205.06240 [gr-qc]].


\bibitem{Zilhao:2012gp}
M.~Zilhao, V.~Cardoso, C.~Herdeiro, L.~Lehner and U.~Sperhake,
``Collisions of charged black holes,''
Phys. Rev. D \textbf{85} (2012), 124062
[arXiv:1205.1063 [gr-qc]].



\bibitem{East:2022rqi}
W.~E.~East and F.~Pretorius,
``Binary neutron star mergers in Einstein-scalar-Gauss-Bonnet gravity,''
Phys. Rev. D \textbf{106} (2022) no.10, 104055
[arXiv:2208.09488 [gr-qc]].


\bibitem{Figueras:2021abd}
P.~Figueras and T.~Fran{\c{c}}a,
``Black hole binaries in cubic Horndeski theories,''
Phys. Rev. D \textbf{105} (2022) no.12, 124004
[arXiv:2112.15529 [gr-qc]].


\bibitem{Herdeiro:2018wub}
C.~A.~R.~Herdeiro, E.~Radu, N.~Sanchis-Gual and J.~A.~Font,
``Spontaneous Scalarization of Charged Black Holes,''
Phys. Rev. Lett. \textbf{121} (2018) no.10, 101102
[arXiv:1806.05190 [gr-qc]].





\bibitem{Bahamonde:2022lvh}
S.~Bahamonde, L.~Ducobu and C.~Pfeifer,
``Scalarized black holes in teleparallel gravity,''
JCAP \textbf{04} (2022) no.04, 018
[arXiv:2201.11445 [gr-qc]].



\bibitem{Bahamonde:2022chq}
S.~Bahamonde, D.~D.~Doneva, L.~Ducobu, C.~Pfeifer and S.~S.~Yazadjiev,
``Spontaneous scalarization of black holes in Gauss-Bonnet teleparallel gravity,''
Phys. Rev. D \textbf{107} (2023) no.10, 104013
[arXiv:2212.07653 [gr-qc]].



\bibitem{Gonzalez:2024ifp}
P.~A.~Gonz{\'a}lez, E.~Papantonopoulos, J.~Robledo and Y.~V{\'a}squez,
``Nonlinear scalarization of Schwarzschild black holes in scalar-torsion teleparallel gravity,''
Phys. Rev. D \textbf{111} (2025) no.4, 044064
[arXiv:2407.13557 [gr-qc]].

\bibitem{Bravo-Gaete:2025lgs}
M.~Bravo-Gaete, A.~Cisterna, M.~Hassaine and D.~Kubiznak,
Phys. Lett. B \textbf{868}, 139721 (2025)
[arXiv:2506.04854 [hep-th]].

\bibitem{Bravo-Gaete:2026bdq}
M.~Bravo-Gaete, J.~A.~Herrera-Mendoza, J.~Oliva and X.~Zhang,
Eur. Phys. J. C \textbf{86}, no.5, 502 (2026)
[arXiv:2601.18048 [hep-th]].




\bibitem{DeFelice:2026cse}
A.~De Felice and S.~Tsujikawa,
``Light rings, gravitational lensing, and ISCOs of exotic compact objects in Einstein-scalar-Maxwell theories,''
JCAP \textbf{06} (2026), 002
[arXiv:2602.23657 [gr-qc]].




\bibitem{Xiong:2022ozw}
W.~Xiong, P.~Liu, C.~Niu, C.~Y.~Zhang and B.~Wang,
``Dynamical spontaneous scalarization in Einstein-Maxwell-scalar theory *,''
Chin. Phys. C \textbf{46} (2022) no.9, 095103
[arXiv:2205.07538 [gr-qc]].




\bibitem{Niu:2022zlf}
C.~Niu, W.~Xiong, P.~Liu, C.~Y.~Zhang and B.~Wang,
``Dynamical descalarization in Einstein-Maxwell-scalar theory,''
[arXiv:2209.12117 [gr-qc]].






\bibitem{Jiang:2023yyn}
J.~Y.~Jiang, Q.~Chen, Y.~Liu, Y.~Tian, W.~Xiong, C.~Y.~Zhang and B.~Wang,
``Type I critical dynamical scalarization and descalarization in Einstein-Maxwell-scalar theory,''
Sci. China Phys. Mech. Astron. \textbf{67} (2024) no.2, 220411
[arXiv:2306.10371 [gr-qc]].




\bibitem{Andrade:2021rbd}
T.~Andrade, L.~Areste Salo, J.~C.~Aurrekoetxea, J.~Bamber, K.~Clough, R.~Croft, E.~de Jong, A.~Drew, A.~Duran and P.~G.~Ferreira, \textit{et al.}
``GRChombo: An adaptable numerical relativity code for fundamental physics,''
J. Open Source Softw. \textbf{6} (2021) no.68, 3703
[arXiv:2201.03458 [gr-qc]].



\bibitem{Kaluza:1921tu}
T.~Kaluza,
``Zum Unit{\"a}tsproblem der Physik,''
Sitzungsber. Preuss. Akad. Wiss. Berlin (Math. Phys. ) \textbf{1921} (1921), 966-972
[arXiv:1803.08616 [physics.hist-ph]].



\bibitem{Fernandes:2019rez}
P.~G.~S.~Fernandes, C.~A.~R.~Herdeiro, A.~M.~Pombo, E.~Radu and N.~Sanchis-Gual,
``Spontaneous Scalarisation of Charged Black Holes: Coupling Dependence and Dynamical Features,''
Class. Quant. Grav. \textbf{36} (2019) no.13, 134002
[erratum: Class. Quant. Grav. \textbf{37} (2020) no.4, 049501]
[arXiv:1902.05079 [gr-qc]].




\bibitem{Friedman:2004jr}
J.~L.~Friedman,
``The Cauchy problem on space-times that are not globally hyperbolic,''
[arXiv:gr-qc/0401004 [gr-qc]].


\bibitem{Baumgarte:1998te}
T.~W.~Baumgarte and S.~L.~Shapiro,
``On the numerical integration of Einstein's field equations,''
Phys. Rev. D \textbf{59} (1998), 024007
[arXiv:gr-qc/9810065 [gr-qc]].



\bibitem{Shibata:1995we}
M.~Shibata and T.~Nakamura,
``Evolution of three-dimensional gravitational waves: Harmonic slicing case,''
Phys. Rev. D \textbf{52} (1995), 5428-5444


\bibitem{Bernuzzi:2009ex}
S.~Bernuzzi and D.~Hilditch,
``Constraint violation in free evolution schemes: Comparing BSSNOK with a conformal decomposition of Z4,''
Phys. Rev. D \textbf{81} (2010), 084003
[arXiv:0912.2920 [gr-qc]].



\bibitem{Alic:2011gg}
D.~Alic, C.~Bona-Casas, C.~Bona, L.~Rezzolla and C.~Palenzuela,
``Conformal and covariant formulation of the Z4 system with constraint-violation damping,''
Phys. Rev. D \textbf{85} (2012), 064040
[arXiv:1106.2254 [gr-qc]].


\bibitem{Bona:1994dr}
C.~Bona, J.~Masso, E.~Seidel and J.~Stela,
``A New formalism for numerical relativity,''
Phys. Rev. Lett. \textbf{75} (1995), 600-603
[arXiv:gr-qc/9412071 [gr-qc]].



\bibitem{Alcubierre:2002kk}
M.~Alcubierre, B.~Bruegmann, P.~Diener, M.~Koppitz, D.~Pollney, E.~Seidel and R.~Takahashi,
``Gauge conditions for long term numerical black hole evolutions without excision,''
Phys. Rev. D \textbf{67} (2003), 084023
[arXiv:gr-qc/0206072 [gr-qc]].


\bibitem{Berger}
M.J. Berger, P. Colella,
Local adaptive mesh refinement for shock hydrodynamics,
Journal of Computational Physics,
Volume 82, Issue 1,
1989,
Pages 64-84,
ISSN 0021-9991.


\bibitem{Kreiss}
Heinz-Otto Kreiss, ``Methods for the approximate solution of time dependent problems,'' 1973. url: https://api.semanticscholar.org/CorpusID:118627871.



\bibitem{Franca:2023bed}
T.~Fran{\c{c}}a,
``Binary Black Holes in Modified Gravity,''
[arXiv:2308.12037 [gr-qc]].



\bibitem{Dreyer:2002mx}
O.~Dreyer, B.~Krishnan, D.~Shoemaker and E.~Schnetter,
``Introduction to isolated horizons in numerical relativity,''
Phys. Rev. D \textbf{67} (2003), 024018
[arXiv:gr-qc/0206008 [gr-qc]].


\bibitem{Baker:2001sf}
J.~G.~Baker, M.~Campanelli and C.~O.~Lousto,
``The Lazarus project: A Pragmatic approach to binary black hole evolutions,''
Phys. Rev. D \textbf{65} (2002), 044001
[arXiv:gr-qc/0104063 [gr-qc]].



\bibitem{Ansorg:2004ds}
M.~Ansorg, B.~Bruegmann and W.~Tichy,
``A Single-domain spectral method for black hole puncture data,''
Phys. Rev. D \textbf{70} (2004), 064011
[arXiv:gr-qc/0404056 [gr-qc]].





\bibitem{Bozzola:2019aaw}
G.~Bozzola and V.~Paschalidis,
``Initial data for general relativistic simulations of multiple electrically charged black holes with linear and angular momenta,''
Phys. Rev. D \textbf{99} (2019) no.10, 104044
[arXiv:1903.01036 [gr-qc]].




\bibitem{Estabrook:1973ue}
F.~Estabrook, H.~Wahlquist, S.~Christensen, B.~DeWitt, L.~Smarr and E.~Tsiang,
Phys. Rev. D \textbf{7} (1973), 2814-2817



\bibitem{York:1973ia}
J.~W.~York, Jr.,
``Conformatlly invariant orthogonal decomposition of symmetric tensors on Riemannian manifolds and the initial value problem of general relativity,''
J. Math. Phys. \textbf{14} (1973), 456-464



\bibitem{Bowen:1980yu}
J.~M.~Bowen and J.~W.~York, Jr.,
``Time asymmetric initial data for black holes and black hole collisions,''
Phys. Rev. D \textbf{21} (1980), 2047-2056




\bibitem{Tichy:2003zg}
W.~Tichy, B.~Bruegmann and P.~Laguna,
``Gauge conditions for binary black hole puncture data based on an approximate helical Killing vector,''
Phys. Rev. D \textbf{68} (2003), 064008
[arXiv:gr-qc/0306020 [gr-qc]].



\bibitem{Zilhao:2013nda}
M.~Zilh{\~a}o, V.~Cardoso, C.~Herdeiro, L.~Lehner and U.~Sperhake,
``Collisions of oppositely charged black holes,''
Phys. Rev. D \textbf{89} (2014) no.4, 044008
[arXiv:1311.6483 [gr-qc]].



\end{thebibliography}
\end{document}